\documentclass[twocolumn,showpacs,preprintnumbers,amsmath,amssymb,aps,pre,superscriptaddress]{revtex4-2}

\usepackage{graphicx}
\usepackage{dcolumn}
\usepackage{bm}
\usepackage{tikz}

\begin{document}


\title{A New Spin on Color Quantization}

\author{Samy Lakhal}
 \affiliation{LadHyX, {UMR} CNRS 7646, Ecole Polytechnique, 91128 Palaiseau Cedex, France}
  \affiliation{Institut Jean Le Rond d’Alembert, UMR CNRS 7190, Sorbonne Université, 75005 Paris, France}
   \affiliation{Chair of EconophysiX \& Complex Systems,  Ecole Polytechnique, 91128 Palaiseau Cedex, France}
 \author{Alexandre Darmon}
  \affiliation{Art in Research, 33 rue Censier, 75005 Paris, France}
\author{Michael Benzaquen}
 \email{michael.benzaquen@polytechnique.edu}
 \affiliation{LadHyX, {UMR} CNRS 7646, Ecole Polytechnique, 91128 Palaiseau Cedex, France}
  \affiliation{Chair of EconophysiX \& Complex Systems,  Ecole Polytechnique, 91128 Palaiseau Cedex, France}
 \affiliation{Capital Fund Management, 23 rue de l'Université, 75007 Paris, France\smallskip}

\date{\today}

\begin{abstract}
We address the problem of image color quantization using a Maximum Entropy based approach. Focusing on pixel mapping we argue that adding thermal noise to the system yields better visual impressions than that obtained from a simple energy minimization. To quantify this observation, we introduce the coarse-grained quantization error, and seek the optimal temperature which minimizes this new observable. By comparing images with different structural properties, we show that the optimal temperature is a good proxy for complexity at different scales. {{Noting} that the convoluted error is a key observable, we directly minimize it using a Monte Carlo algorithm to generate a new series of quantized images.} Adopting an original approach based on the informativity of finite size samples, we are able to determine the optimal convolution parameter leading to the best visuals. {Finally, we test the robustness of our method against changes in image type, color palette and convolution kernel.}
\end{abstract}

\maketitle





\section*{Introduction}

 In physics, many problems can be formulated in terms of energy minimization. However, for complex systems with a large number of degrees of freedom, analytical minimizers are often difficult to find, and the ground state is seldom representative of the true physical state of the system at hand.
To overcome such issues, a classical method in statistical thermodynamics is to slightly relax the energy minimization constraint and introduce a probabilistic model relying on entropy maximization~\cite{jaynes_information_1957,jaynes_information_1957-1}. Such an approach has allowed for the exploration of suboptimal solutions with thermal noise and led to the emergence of historical results on phase transitions, e.g. for Ising models \cite{ising_beitrag_1925,onsager_crystal_1944}. This method has since been popularized in various fields, for example in biology for inference problems \cite{de_martino_introduction_2018} or in computer science for classifiers (see e.g. the softmax function \cite{bridle_training_1989}) and annealing procedures \cite{kirkpatrick_optimization_1983}. 

Let us now consider the problem of field quantization and its application for images, that is color quantization.
It consists in choosing a set of authorized states called the color palette and then projecting each pixel of the original image onto this palette. This method is naturally very relevant for compression and other problems involving digital image processing. The first step consists in finding the most convenient color palette from the original histogram using  thresholding levels  \cite{heckbert_color_1982,gervautz_simple_1988} or clustering methods \cite{balasubramanian_sequential_1994,dekker_kohonen_1994,abernathy_incremental_2022,huang_efficient_2021}. The second step, {called \textit{pixel mapping},
is usually achieved by performing a simple nearest color procedure. Stochastic mapping -- or \textit{dithering procedures} --  are also used  to reduce threshold artifacts and improve the overall visual quality of the quantized image~\cite{ulichney_digital_1987,orchard_color_1991,puzicha_spatial_2000}.}

Here, we shall focus on the pixel mapping step by constraining the color palette \textit{a priori}. { The aim of the present study is to use seasoned statistical physics tools and novel information theory methods - which come with a high degree of physical interpretability - to tackle the color quantization problem.} We adopt a field theory approach, based on the exploration of simple observables at and around optimality using thermal noise.   { In Section I} we  provide a method to explore new solutions to the quantization problem with a Maximum Entropy based approach. { In Section II} we show that, in the specific case of color quantization and with the simplest error measure, this method yields surprisingly good visuals when varying the temperature of the system. {To quantify this observation, { and inspired by \cite{puzicha_spatial_2000}, we use} the convoluted error to look for optimal thermal noise levels regarding the overall image quality.} { In Section III} we  confront the results for images with different structural characteristics, and show that the optimal temperatures are a good proxy for image complexity. 
{ In Section IV}, {  guided by  \cite{puzicha_spatial_2000}}, we implement a Monte Carlo algorithm to directly minimize the convoluted error, and, following a novel { method} on the informativity of finite size samples~\cite{cubero_minimum_2018, cubero_multiscale_2020, haimovici_criticality_2015}, we determine the convolution parameter providing the best visuals {in Section V}. {Finally, in Section VI we illustrate how the method can be extended to other types of images, different target color palettes and alternative convolution kernels.}

\section{\label{sec:level2}State quantization}
We consider a field $h(\mathbf{r}) \subset F$  and want to build its optimal quantized version $ \hat h(\mathbf{r})  \subset \hat F$ where $\hat F$ is a subset of $F$. To do so, one usually minimizes a loss function between $\hat h$ and $h$. A first natural choice for the loss function is a site-wise measure of the quantization error $\mathcal{L}(h,\hat h) = \sum_{\mathbf{r}} \mathcal{L}_{F}(h(\mathbf{r}),\hat h(\mathbf{r}))$ where $\mathcal{L}_F$ is the {loss for each site}, usually Euclidean or logical. With such a definition, the field $\hat h^{\ast}$ minimizing $\mathcal L(h,\hat h)$ is simply obtained by replacing each original {data} with the closest {state} in $\hat F$. Note that this is what commonly happens during the sampling of a continuous signal with an instrument, such as a camera projecting colors in the RGB space \cite{rowlands_color_2020}. 
There are cases where this simple quantization process  leads to unsatisfying $\hat{h}^{\ast}$ fields deviating too much from the original data. For example, quantizing a {continuous white noise} on a grid with a threshold level artificially generates correlated samples of the site percolation problem~\cite{voss_fractal_1984,saberi_recent_2015,shante_introduction_1971}.

To engineer more relevant loss measures, our idea is  to explore suboptimal configurations around $\hat h^{\ast}$ using the Maximum Entropy approach mentioned above~\cite{jaynes_information_1957}. This allows one to define the most agnostic -- that is the most entropic -- classes of distributions with given constraints, such as normalization.
In the context of state quantization we look for the distributions $(\mathcal P)$ over quantized fields that maximize the following functional:
\begin{multline}
    J(\mathcal P) = S(\mathcal P) -  \mu [\sum_{\hat h}\mathcal P(\hat h) - 1]\\ 
    - \lambda \sum_{\hat h} \mathcal P(\hat h) \left[ \mathcal L(h,\hat h) - \mathcal L(h,\hat h^{\ast})\right],
    \label{equation:MED}
\end{multline}
where the first term $S(\mathcal P) = -\sum_{\hat h} \mathcal P(\hat h) \log \mathcal P(\hat h)$ is the distribution entropy, the second term is the normalization constraint, the last one a constraint on quantization error, and $\mu$ and $\lambda$ their respective Lagrange multipliers. Differentiating Eq.~\eqref{equation:MED} with respect to $\mathcal P(\hat h)$ and  $\mu$ allows to enforce normalization while leaving $\lambda$ as a free parameter. This leads to: 
\begin{equation}
    \mathcal P_{h,\lambda}(\hat h) = \frac{1}{Z_{h,\lambda}}e^{-\lambda \mathcal L(h,\hat h)},
    \label{equation:3}
\end{equation}
where $Z_{h,\lambda}$ is the partition function. By setting $T := 1/\lambda$, we recover a Boltzmann-like distribution where the loss function plays the role of the energy. In extreme cases such as $T \to 0$ and $T \to + \infty $, we respectively recover the Dirac delta distribution centred on $\hat h^{\ast}$ and the uniform distribution. As in the study of any Hamiltonian system, increasing the temperature softens the energy minimization constraint and is therefore the opportunity to test other basic observables.

 \section{ Pixel mapping}\label{sec2}

We now apply the above formalism in the classical image processing problem of grayscale quantization. The goal is to reduce the amount of shades taken by the pixels in an image, usually described with 256 levels. In this context, $h$ and $\hat h$ respectively correspond to the original and reduced images, while $F$ and $\hat F$ respectively correspond to the initial and quantized sets of grayscale levels. 
To define the loss function we use the naive Squared Error, obtained from squaring the Euclidean distance:
\begin{equation}
    \mathcal L(h,\hat h) = \|h-\hat h\|_2^2=
    \sum_{\mathbf{r}} \big[h(\mathbf{r})-\hat h(\mathbf{r})\big]^2.
    \label{eq:norm2}
\end{equation}
This loss function being pixel-wise separable, so is the corresponding distribution  (Eq.~\eqref{equation:3}):
\begin{equation}
    \mathcal P_{h,T}(\hat h) = \prod_{\mathbf r} p_{h(\mathbf r),T}(\hat h(\mathbf{r})) = \prod_{\mathbf{r}}\frac{1}{z_{h(\mathbf{r}),T}} e^{-[h(\mathbf{r}) - \hat h(\mathbf{r})]^2/T},
    \label{eq:marginalized}
\end{equation}
where $z_{h(\mathbf{r}),T}$ is the partition function of the marginal distribution $p_{h(\mathbf{r}),T}$.

\begin{figure}
    \centering
    \includegraphics[width = \linewidth]{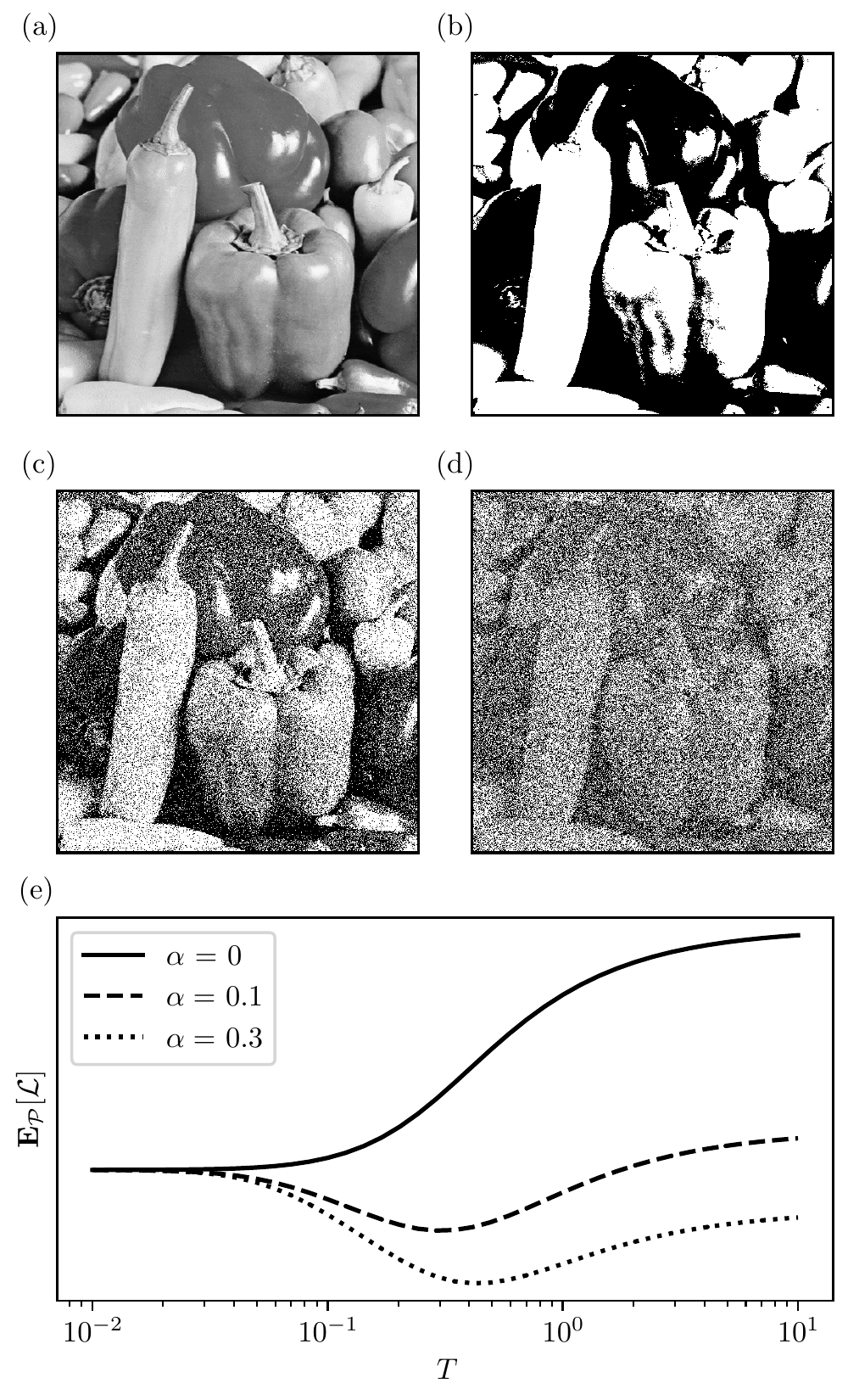}
    \caption{Influence of thermal noise on color quantization. (a) Original benchmark image. {(b-d)} Quantized versions of the original image generated at low $(T\simeq 0)$, intermediate $(T=0.3)$ \& high $(T=1)$ temperatures using Eq.~\eqref{eq:marginalized}.
    (e) Evolution of the rescaled Mean Convoluted Squared Error between the original and quantized images (Eq.~\eqref{eq:ConvolutedDistance}) with temperature for different values of the convolution parameter~$\alpha$ (Eq.~\eqref{eq:kernelFourier}).}
    \label{fig:1}
\end{figure}

To test our method, we use a classical benchmark image {taken from the USC-SIPI database \cite{noauthor_sipi_nodate2}}, shown in Fig.~\ref{fig:1}(a), and we sample quantized versions by using the marginal distributions defined in Eq.~\eqref{eq:marginalized}. The authorized colors are chosen as black and white, meaning that $\hat F = \{0,255\}$.
Figure~\ref{fig:1}(b) was generated at $T\to 0$, equivalent to the naive minimization $\hat h^{\ast}$ presented in Sec.~\ref{sec:level2}, where any texture in the $[0,127]$ or $[128,255]$ intervals is simply replaced by black or white pixels respectively. The image, although still recognizable, displays thresholding artifacts such as contouring effects for shaded textures and suppresses a vast amount of details. In Fig.~\ref{fig:1}(c), we introduce thermal noise with intermediate temperature $T=0.3$, leading to a more interesting visual. Parts of the lighter and darker shades are reconstructed and other details like contours now accurately correspond to the real physical features of the objects, and no longer to fluctuations around the threshold value. In Fig.~\ref{fig:1}(d), one can see that a higher thermal noise level no longer represents a positive contribution, as one excessively randomizes the pixel attribution rule.

Figure~\ref{fig:1}(c) thus interestingly appears to be a better quantized version than Fig.~\ref{fig:1}(b)\&(d), especially when looking at it from a distance. Precisely, taking a step back has the effect of coarse-graining/convoluting the image and erasing the small-scale fluctuations created by thermal noise. To quantify such an observation, we define the Convoluted Squared Error $\mathcal{L}_\theta$ comparing the original and quantized fields after a convolution through a given kernel $\theta$:
\begin{equation}
    \mathcal{L}_\theta(h,\hat h) = \|(h-\hat h)\circledast \theta \|_2^2.
    \label{eq:ConvolutedDistance}
\end{equation}
Among the many classes of kernels commonly used in image processing, we choose a power-law kernel of the form  $\theta_{\gamma}(\mathbf{r}) \propto \|\mathbf{r} \|^{-\gamma}$, for the sake of physical interpretability and mathematical tractability. 
Its Fourier transform is also a power-law: \begin{equation}
\tilde{\theta}_{\alpha}(\mathbf k) \propto \|\mathbf{k} \|^{-\alpha},  \label{eq:kernelFourier}
\end{equation}
with $\alpha = d/2 - \gamma$, where $d=2$ is the space dimension. 

When $\alpha = 0$, the kernel in the direct space is  narrow and leaves the image invariant. As $\alpha$ increases, the convolution operation replaces each pixel value with its local average of the field.  Figure~\ref{fig:1}(e) displays the Mean Convoluted Squared Error $\mathbb{E}_{\mathcal{P}_{h,T}} [\mathcal{L}_{\theta}]$ as a function of temperature (see {Appendix~}\ref{Appendix:MSE} for the details of the computation of $\mathbb{E}_{\mathcal{P}_{h,T}}[\mathcal{L}_{\theta}]$).
As expected, the unconvoluted Squared Error ($\alpha = 0$, solid line) increases monotonously with $T$. For higher values of $\alpha$, however, a local minimum appears at a finite temperature $T^\star_{\alpha}$. 
 Note that Fig.~\ref{fig:1}(c) was generated with a noise level $T$ close to the minima displayed in Fig.~\ref{fig:1}(e), { thereby confirming  that the Mean Convoluted Squared Error is a relevant observable for the color quantization problem.}

\begin{figure}[h]
    \centering
    \includegraphics[width = \linewidth]{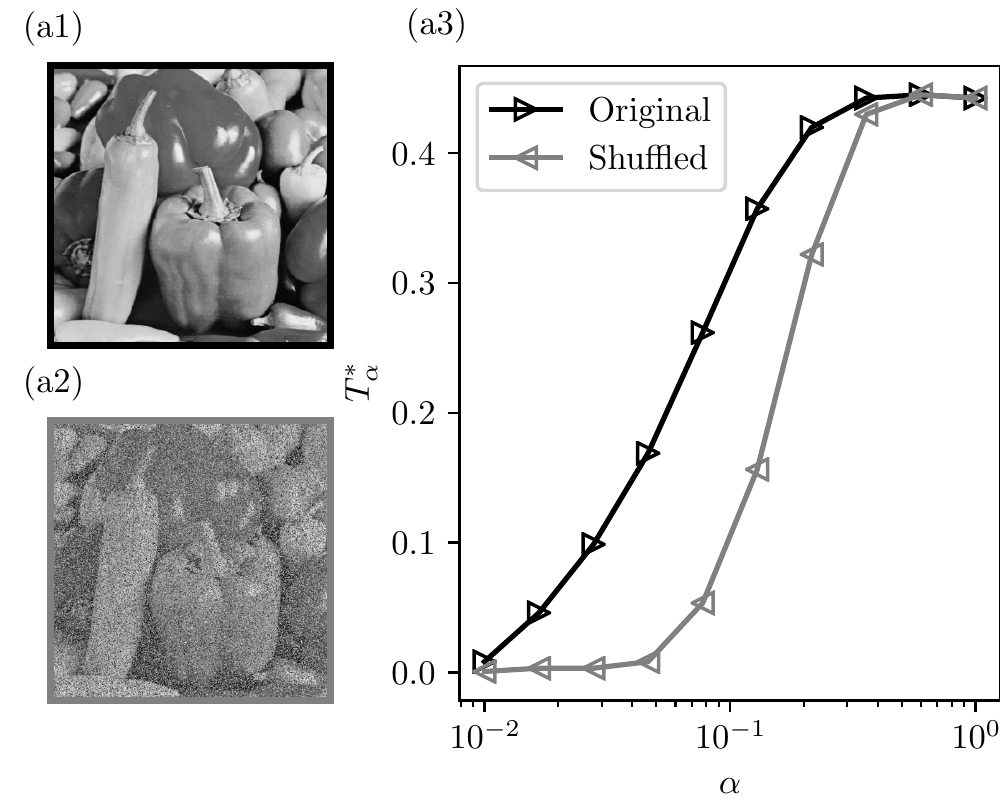}
    \includegraphics[width = \linewidth]{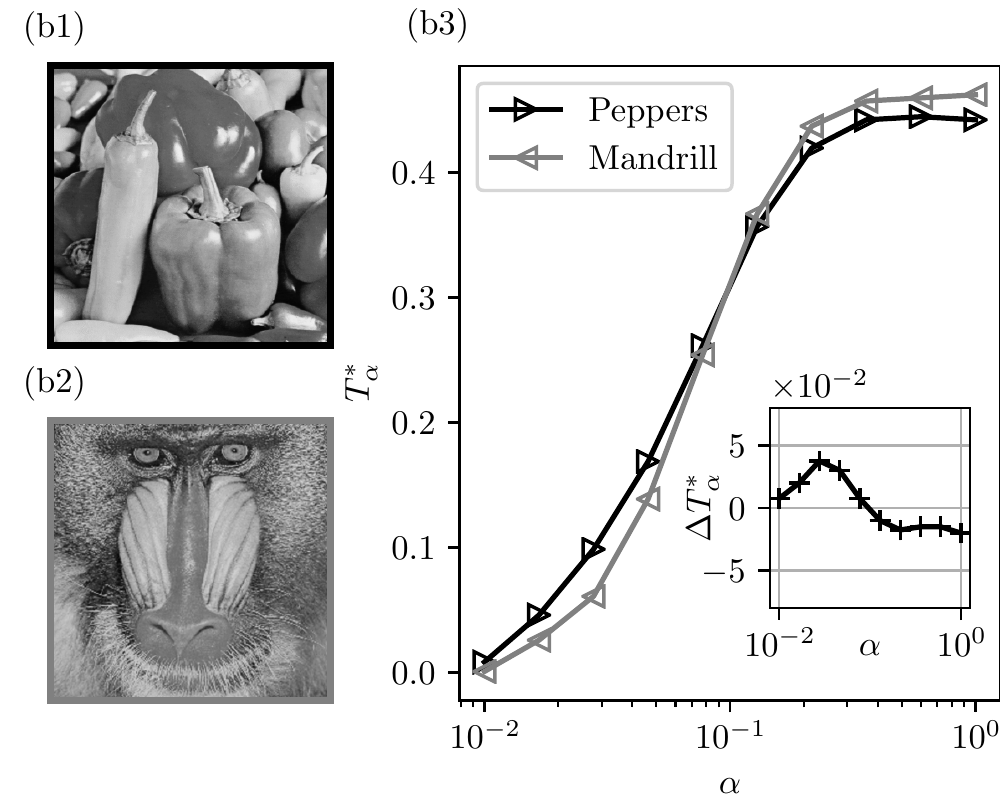}
    \caption{Evolution of the optimal temperature $T^\star_\alpha$ with the convolution parameter $\alpha$ for different images. (a) Comparison between the original image and a shuffled version generated with a randomizing procedure. (b) Comparison between two benchmark images (Peppers \& Mandrill) displaying structural features at different scales.}
    \label{fig:2}
\end{figure}

\section{Visual complexity}

The optimal temperature $T^\star_\alpha$ naturally depends on the color histogram of the original image, but also on the spatial arrangement of its pixels.
To quantify this last statement, we compare our benchmark image with its transformation through a  histogram-invariant operation. We use  
{{a shuffling procedure that randomly selects two pixels and switches their position. The procedure is then repeated until the number of switching operations is equal to the number of pixels in the image, that is $512^2$. }} In Figs.~\ref{fig:2}(a1) and (a2), we respectively display the original image and its shuffled version, and we plot the temperatures $T^\star_\alpha$ as function of $\alpha$ in Fig.~\ref{fig:2}(a3). {{Note that the shuffled image is still recognizable as some pixels are never selected by the procedure}}. For low values of $\alpha$, the kernel is too narrow and the convolution has almost no effect, unsurprisingly leading to $T^\star_\alpha = 0$ for both images. Then, both temperatures monotonously increase with $\alpha$ until they meet at a plateau where the evolution is independent of the spatial distribution of pixels. Interestingly, we observe that $T^\star_\alpha$ is systematically higher for the original image, meaning that the shuffling procedure has strongly affected its structural properties. Indeed, natural images such as the peppers present intelligible patterns and strong spatial regularities, far from the random and uncorrelated rearrangement that the shuffling procedure creates. This behaviour is somehow reminiscent of several classical physical systems such as the Random Field Ising Model (RFIM) for which irregularities lower the critical temperature \cite{fishman_random_1979,belanger_random_1991} (see {Appendix~}\ref{Appendix:Ising} for more details on the link between the Convoluted Squared Error and the RFIM Hamiltonian). 

In Fig.~\ref{fig:2}(b), we test another approach by comparing the evolution of $T^\star_\alpha$ with $\alpha$ for two different benchmark images, each presenting interesting visual features at different scales. Figure~\ref{fig:2}(b1) presents less small scale features than Fig.~\ref{fig:2}(b2), resulting in a higher optimal temperature for low values of $\alpha$, see Fig.~\ref{fig:2}(b3) (the inset shows the difference of temperatures $\Delta T^\star_\alpha$ between the two images). This tendency reverses at higher scales. 
Using this procedure we compared a number of other types of natural images (forests, fields, buildings, landscapes), as well as classes of simple abstract textures like those presented in~\cite{lakhal_beauty_2020}, with the same conclusions. This supports the intuition that the temperatures $T^\star_\alpha$ may be used as measures of multi-scale visual complexities and as such, consistent input features for aesthetic assessment algorithms~\cite{valenzise_advances_2022,yuan_opinion-unaware_2020,maitre_esthetique_2022,bagrov_multiscale_2020,mccormack_enigma_2021}.

\begin{figure}
    \centering
    \includegraphics[width = \linewidth]{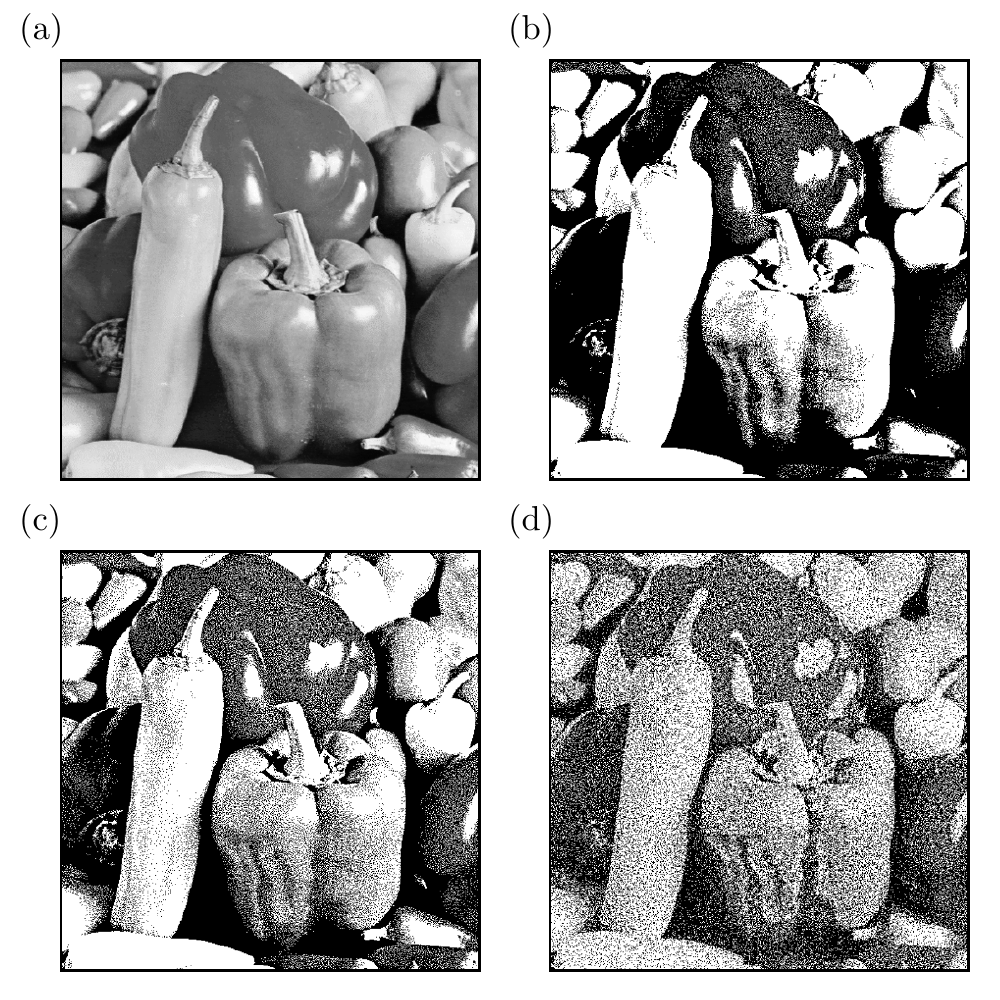}
    \caption{Monte Carlo image generation. (a) Original benchmark image. (b-d) Monte Carlo simulations for $\alpha=0.02$, $\alpha=0.05$, $\alpha=0.5$ respectively. Images were initialized with $\hat h^{\ast}$ and the simulation ran until the loss function reached stability.}
    \label{fig:3}
\end{figure}

\section{\label{sec:level3} Monte Carlo image generation}

As argued above, the Convoluted Squared Error $\mathcal{L}_\theta$ in Eq.~\eqref{eq:ConvolutedDistance} is a highly relevant observable for the color quantization problem. As such, it seems reasonable to use it as our effective energy function. An idea, see e.g.~\cite{puzicha_spatial_2000}, is to directly generate images minimizing this error function using a Monte Carlo algorithm. This algorithmic approach is necessary as there is no explicit minimum of~$\mathcal L_{\theta}( h,\cdot)$. { Let us stress that, while Monte Carlo approaches and other dithering methods have historically contributed to the image quantization problem, the last two decades were marked by more evolved developments, using adaptive kernels \cite{huang_efficient_2016} and clustering algorithms \cite{thompson_fast_2020,frackiewicz_fast_2019}. Here, for the sake of physical interpretation, we focus our attention on the classic Monte Carlo approach as it contains the minimal ingredients to tackle this problem.}

Here we implement a simplified Monte Carlo algorithm where we use the power-law kernel defined in Eq.~\eqref{eq:kernelFourier} (rather than the Gaussian window with simulated annealing procedures used in~\cite{puzicha_spatial_2000}).
Images were initialized with the solution $\hat{h}^{\ast}$, and the algorithm ran until  convergence of the loss functions. For algorithmic efficiency, we use the Parsival equality on Eq.~\eqref{eq:ConvolutedDistance} to obtain $\mathcal{L}_{\theta}(h,\hat h) = \| \mathcal F [h - \hat{h}](\mathbf k)\cdot k^{-2 \alpha}\|_2^2  $, with $\mathcal F $ the Fourier transform
  calculated using the FFT algorithm (thereby assuming periodic boundary conditions). Figure~\ref{fig:3}(a) displays again the original image for reference, and {Figs.~\ref{fig:3}(b-d)} the images corresponding respectively to $\alpha=0.02$, $\alpha=0.05$, $\alpha=0.5$. 
For Fig.~\ref{fig:3}(b), the kernel function is narrow: small-scale details like the shadows on the peppers are faithfully reproduced, while leaving large areas of uniform color. Increasing $\alpha$ helps removing the latter artefact and improves the overall visual impression when looked at from a distance{, see Fig.~\ref{fig:3}(c)}. However, too large convolution windows yield images lacking small-scale accuracy, see Fig.~\ref{fig:3}(d). A compromise shall thus be found in order to generate the most faithful quantized image, that is the optimal  $\alpha$ providing the best trade-off throughout different scales.

 \section{\label{sec:optimal}Optimal pixel mapping}

{In classical dithering algorithms, calibration is usually achieved using perception-related functionals, such as the structural similarity index measure (SSIM), that mimick human visual quality assessment~\cite{ramella_evaluation_2021,wang2004image}. Although very good results can be obtained with such metrics~\cite{jaques_novel_2022}, the method lacks physical interpretability. Here, we wish to provide a sounder entropy-based approach that captures the ability of images to retain information during compression, as shown in recent visual appreciation experiments~\cite{lakhal_beauty_2020}.}

To find the most suitable parameter $\alpha$, we use a recent information theory based approach~\cite{cubero_minimum_2018} where the spatial resolution becomes a tunable parameter of the system. 
{
This method has  proven to be highly efficient in other fields such as neuroscience~\cite{cubero_multiscale_2020},   finance~\cite{haimovici_criticality_2015},  deep learning~\cite{song2018resolution},
or language models~\cite{marsili2022quantifying}.}

First, one defines the image as a sample $(\mathbf r,\hat h(\mathbf r))_{\mathbf r}$ of pixels. Each tuple being unique,  this description is far from being the most efficient in the context of information theory. Indeed, information could for example be saved by accounting for neighboring pixels with the same color value. One thus modifies the spatial resolution  by considering a new sample $(G^{\ell}(\mathbf r),\hat h(\mathbf r))_{\mathbf r}$ where $G^{\ell}(\mathbf r)$ is the new position on a coarser grid of step ${\ell\in \{1,...,L\}}$, with $L$ the size of the original image. Several tuples can now be in the same state $s$ and one can look at their occurrence -- or degeneracy level -- $k$ inside each block. Finally, one  constructs two measures of entropy $\hat H^{\ell}[s]$ and $\hat H^{\ell}[k]$, respectively assessing the heterogeneity in the data and the heterogeneity in the data redundancy. 

\begin{figure}[t!]
    \centering
    \includegraphics[width = \linewidth]{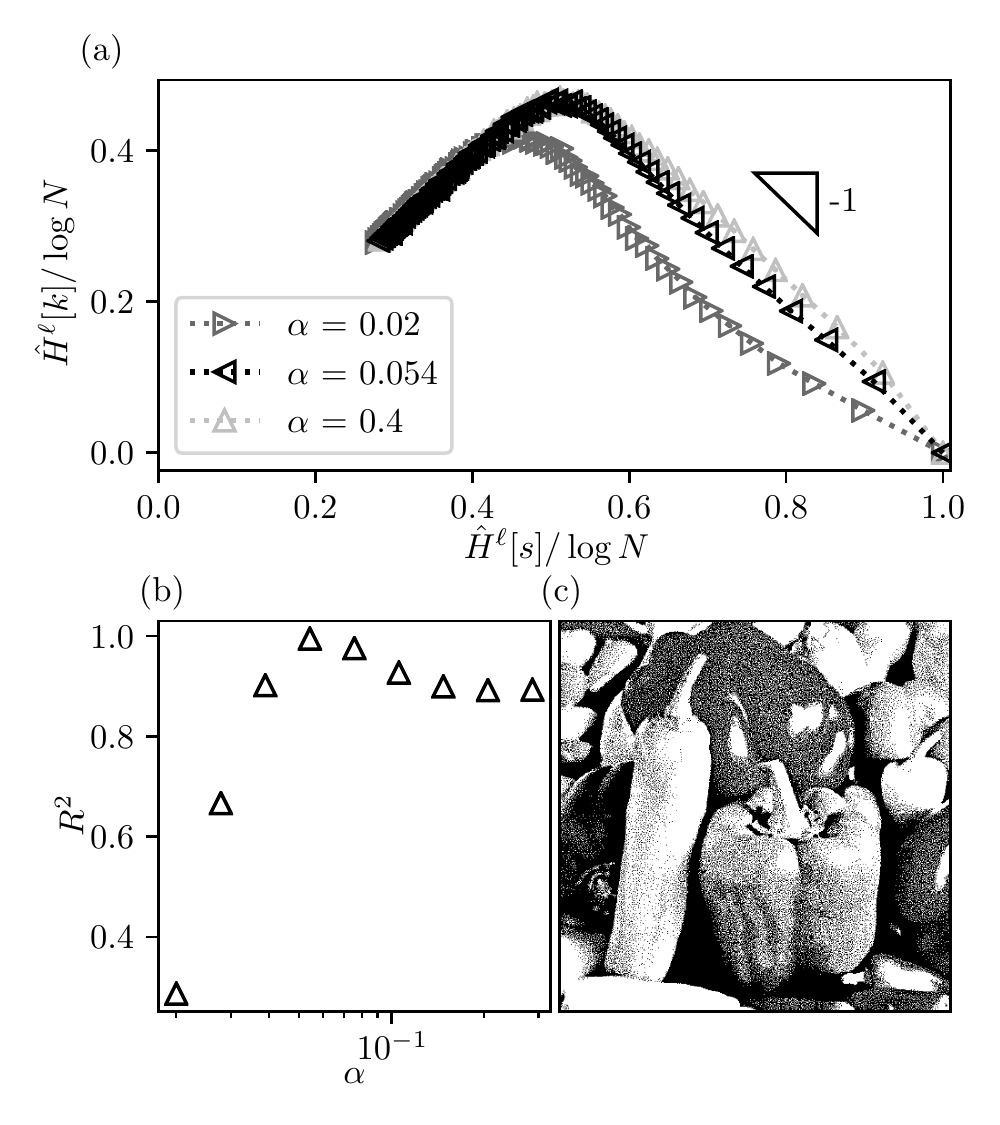}
    \caption{Influence of the convolution parameter $\alpha$ on the compressing regime of color quantized images. (a) Plot of $(H^{\ell}[s],H^{\ell}[k])_{\ell}$ for low, optimal and high convolution parameters $\alpha$. (b) Regression coefficient $R^2$ of the linear fit as function  of $\alpha$. (c) Optimally quantized image, $\alpha = 0.054$.}
    \label{fig:MIR}
\end{figure}

In Fig.~\ref{fig:MIR}(a), we vary $\ell$ and plot $\hat H^{\ell}[k]$ as function of $\hat H^{\ell}[s]$ for images generated with different $\alpha$.  Of most interest to us here is the right part of the graphs, which corresponds to small values of $\ell$, and for which the concavity of the curves is very $\alpha$-dependent.
Indeed, one can show -- see {Appendix~}\ref{Appendix:MIR} -- that the local slope $\mu =\text{d}\hat{H}[k]/\text{d}\hat{H}[s]$  actually corresponds to the {trade-off rate between relevance and resolution as data is compressed}.
The idea is then to choose the convolution parameter such that {this trade-off is as stable as possible across all scales of observations}, meaning that $\mu$ should be close to -1 and as constant as possible when varying $\ell$. In other terms, the right part of the graph should be as linear as possible with slope $-1$. In Fig.~\ref{fig:MIR}(b) we plotted the regression coefficient $R^2$ obtained from the corresponding linear fit and found that it was maximized for an intermediate value {$\alpha = 0.054$}. The corresponding image is shown in Fig.~\ref{fig:MIR}(c) and is indeed a very good visual compromise.

\begin{figure*}
\centering
    \includegraphics[width = .195\textwidth]{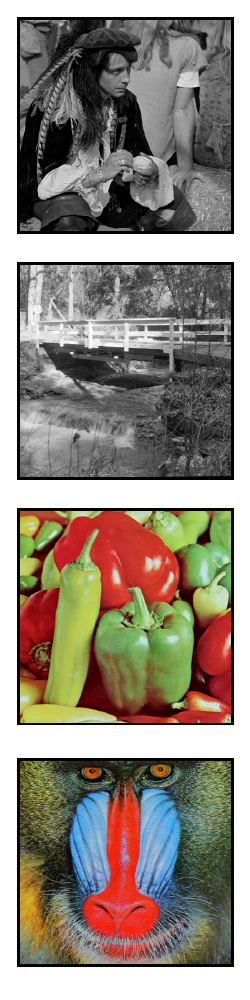}
    \includegraphics[width = .195\textwidth]{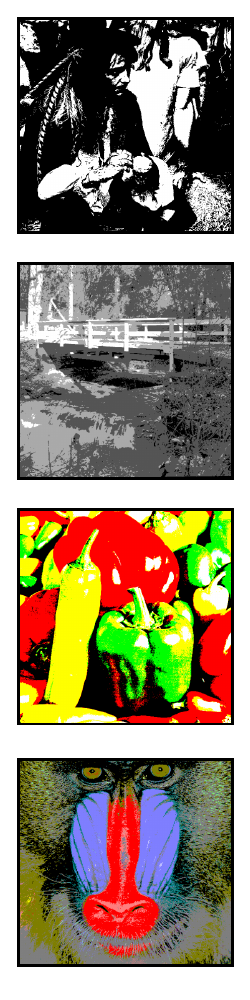}
    \includegraphics[width = .195\textwidth]{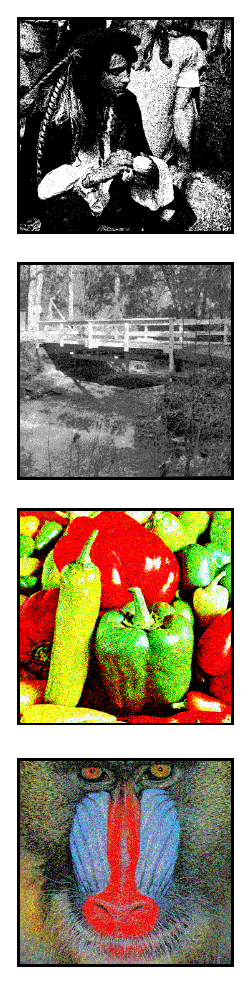}
    \includegraphics[width = .195\textwidth]{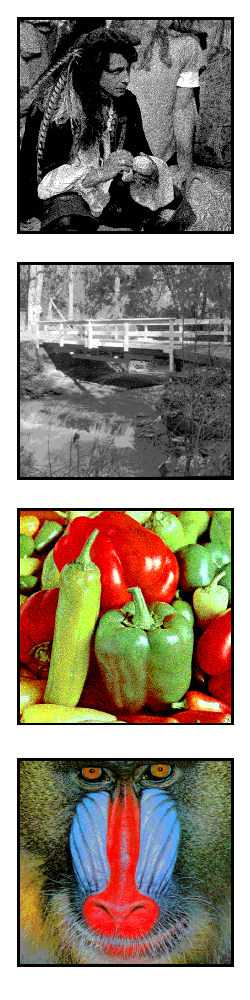}
    \includegraphics[width = .195\textwidth]{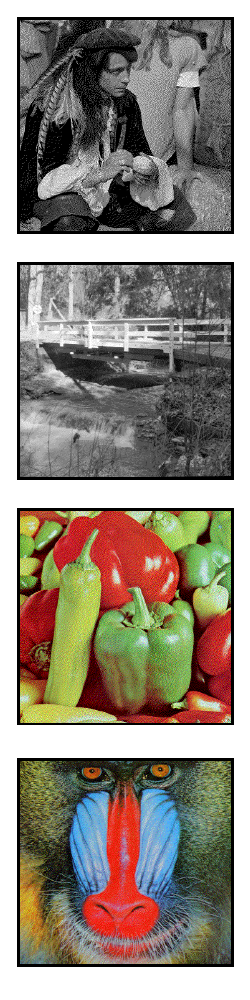}

    \caption{{Extension to different image types and color palettes. From Left to Right: Original image, $T=0$, $T=T^*$, $\sigma = \sigma^*$, and Floyd-Steinberg algorithm. From Top to Bottom: Man onto B\&W, Bridge onto four grayscale levels, Peppers onto eight colors, and Mandrill onto fifteen colors.}}
    \label{fig:panel}
\end{figure*}

\section{\label{sec:}{Beyond Black \& White}}

{
In this section, we illustrate how our method can be extended to other image types, color spaces and convolution kernels. We use four images from the USC-SIPI image database \cite{noauthor_sipi_nodate2} (see Fig.~\ref{fig:panel}, first column). For each image, we choose a different target color palette, namely B\&W for Man pictures (first row), four grayscale levels for Bridge pictures (second row), eight colors (corresponding to the vertices of the RGB cube) for Peppers pictures (third row), and fifteen colors (corresponding to each point of the face-centered cubic arrangement in the RGB cube) for Mandrill pictures (fourth row).}

{First, we consider the $T\simeq0$ case (second column). For each image, as in Sec.~\ref{sec2}, the quantization process suppresses a large amount of details and produces thresholding artifacts such as contouring effects for shaded textures. Then, we increase the temperature and find the optimal $T^\star$ for each quantized image, see in third column. As it was the case for the B\&W Peppers in Sec.~\ref{sec2}, a clear improvement is obtained for each image at $T=T^\star$, with the reconstruction of features such as contours and shades. However, the latter are not yet fully consistent with the original image. The level of noise is still quite high when looking up close. Finally, we generate quantized images with the Monte-Carlo algorithm (fourth column) and test its robustness by using classical Gaussian kernels with standard deviation $\sigma$ instead of power-law kernels. Using the method described in Sec.~\ref{sec:optimal} we find, for all images, values of optimal $\sigma^* \approx 0.5$ pixels. The results are then much more satisfactory as the impression of color saturation and level of noise are significantly reduced. Note that our optimal kernels are slightly narrower than the classical error diffusion matrix kernels historically used in dithering algorithms, such as Floyd-Steinberg (FS) \cite{Floyd:1976:AAS} or Jarvis-Judice-Ninke (JJN) \cite{Jarvis_1976}. The latter provide excellent results but their implementation does not allow for the tuning of the kernel width. For the sake of visual comparison, we have implemented the FS algorithm (fifth column) with indeed a slightly larger error dispersion (equivalent to $\sigma \simeq 0.9$ pixels). While the local FS algorithm takes a pragmatic and very efficient turn, our method is more global and above all allows for physical interpretability.}

\section{Conclusion}

Let us summarize what we have achieved. { In the context of color quantization, we confirmed that the naive approach consisting in a simple error minimization does not generally bring satisfactory visuals.} To overcome this issue, we introduced thermal noise through a Maximal Entropy based approach and generated quantized images with  more interesting visuals. To quantify this visual impression, { and guided by \cite{puzicha_spatial_2000}}, we introduced the Convoluted Squared Error, which compares the original and quantized fields after a coarse-graining procedure. Introducing convolution allowed us to find the optimal temperatures minimizing the new observable. Interestingly, we found that such temperatures are a good indicator for complexity at different scales. Moreover, having {confirmed} that the Convoluted Squared Error is a highly relevant observable with respect to color mapping, we directly minimized it to generate new images using a Monte Carlo algorithm. In order to find the optimal convolution parameter leading to the best visuals, we used an original approach based on the informativity of finite size samples. {Finally, we proved the robustness of our method against changes in image type, color palette, and convolution kernel.} 

Note that, as mentioned in the introduction, in the present analysis we have chosen the target color palette \textit{ex ante}, {focusing on simple cases, namely, B\&W, 4 grayscale, RGB vertices and face-centered cubic}. It would be interesting to consider extended color palettes where both the dimension and color values are optimized for a given image. The latter most likely depends on the features of the color histogram of the original image, but also their spatial arrangement. 
Future research should also be devoted to considering alternatives to the Euclidean distance {such as perception based cost functions~\cite{huang_efficient_2016}, structural similarity metrics~\cite{wang2004image}, quality indices~\cite{frackiewicz_fast_2019}}, {distances including transport terms~\cite{ruschendorf1985wasserstein} or edge detection \cite{huang_efficient_2016}}, which take into account \textit{a priori} the local arrangement of the pixels.  


\section*{Acknowledgments}

We warmly thank R. Benichou and J.-P. Bouchaud who contributed to the early stages of this work, as well as M. Marsili and I. Mastromatteo whose input on seeking the optimal $\alpha$ was crucial to conclude our analysis.   We also thank C. Aubrun, J. Garnier-Brun, P. Mergny and A. Pammi for fruitful discussions. This research was conducted within the Econophysics \& Complex Systems Research Chair, under the aegis of the Fondation du Risque, the Fondation de l’Ecole polytechnique, the Ecole polytechnique and Capital Fund Management. 


\bibliographystyle{apsrev4-1}
\nocite{apsrev41Control}
\setlength{\bibsep}{1.8pt plus 0.3ex}
\bibliography{apssamp}

\appendix

\section{Mean (Convoluted) Squared Error}\label{Appendix:MSE}
Here we derive the expression of the Mean Convoluted Squared Error (Fig.~\ref{fig:1}(e)).
Without convolution, the expectation of the loss function reads:
\begin{equation}
\begin{split}
    \mathbb{E}_{\mathcal P} \left[ \mathcal L(h,\hat h)\right] & = \mathbb{E}_{\mathcal P} \left[ \|\hat h -h \|_2^2 \right]\\
    & = \sum_{\mathbf{r}} \mathbb{E}_{\mathcal P} \left[\left(\hat h(\mathbf{r}) -h(\mathbf{r})\right)^2 \right],\\
\end{split}
\label{EPL}
\end{equation}
where we have used the linearity of $\mathbb{E}_{\mathcal P}$.
Each term inside the expectation depends on the marginal variable $\hat h(\mathbf{r})$ and can be easily computed site-by-site, such that Eq.~\eqref{EPL} can be rewritten as:
\begin{equation}
\begin{split}
    \mathbb{E}_{\mathcal P} \left[ \mathcal L(h,\hat h)\right] & = \sum_{\mathbf{r}} \mathbb{E}_{p_{h(\textbf r )}} \left[\left(\hat h(\mathbf{r}) -h(\mathbf{r})\right)^2 \right].
\end{split}
\label{EPL2}
\end{equation}
Applying the same ideas to the Mean Convoluted Squared Error yields:
\begin{eqnarray}
    \mathbb{E}_{\mathcal P} \left[ \mathcal L_\theta(\hat h,h) \right] & =& \mathbb{E}_{\mathcal P} \left[ \|(\hat h -h)\circledast\theta \|_2^2 \right]\nonumber\\
    & =& \sum_{\mathbf{r}} \mathbb{E}_{\mathcal P} \left[\big((\hat h -h)\circledast\theta(\mathbf{r})\big)^2 \right].
\end{eqnarray}
Developing the square  leads to: $\mathbb{E}_{\mathcal P} [ \mathcal L_\theta(\hat h,h) ] =$
\begin{eqnarray}
\sum_{\mathbf{r},\mathbf{r_1},\mathbf{r_2}} \theta(\mathbf{r} - \mathbf{r_1})\theta(\mathbf{r} - \mathbf{r_2}) \times\qquad \quad\qquad \qquad \quad \quad\nonumber\\
\mathbb{E}_{\mathcal P} \left[ \left(\hat h(\mathbf{r_1}) -h(\mathbf{r_1})\right)\left(\hat h(\mathbf{r_2}) -h(\mathbf{r_2})\right)\right].\ 
\end{eqnarray}
The marginal distributions being independent, non-diagonal second order terms (with $\mathbf{r_1} \neq \mathbf{r_2}$) inside the expectation can be written as products of order 1 expectations. Rewriting the diagonal in terms of squared expectations leads to the following expression:
\begin{multline}
\mathbb{E}_{\mathcal P} \left[ \mathcal L_\theta(\hat h,h) \right]  =\|\mathbb{E}_{\mathcal P}[\hat h - h]\circledast\theta \|_2^2 \\
+ \left(\mathbb{E}_{\mathcal P} \left[ \mathcal L(h,\hat h)\right]- \| \mathbb{E}_{\mathcal P}[ \hat h -h ]\|_2^2 \right)\|\theta\|_2^2,
\end{multline}
where $\mathbb{E}_{\mathcal P} [ \mathcal L(h,\hat h)]$ is the previously defined Mean Squared Error in Eq.~\eqref{EPL2}, and $\mathbb{E}_{\mathcal P}[\hat h - h]$ is the field computed using the marginal distributions on each site.

\section{Analogy with the RFIM} \label{Appendix:Ising}

Here we show that the Convoluted Squared Error can be rewritten as the Hamiltonian of a Random Field Ising Model (RFIM). 
Developping the norm in Eq.~\eqref{eq:ConvolutedDistance}, one obtains:
\begin{equation}
    \mathcal L_{\theta}(\hat{h},h) =\|(\hat{h} - h)\circledast\theta \|_2^2 =\sum_{\mathbf{r}} [(\hat{h}- h)\circledast\theta  (\mathbf{r})]^2.
\end{equation}
Rewriting explicitly the convolution product and defining the interaction matrix $J$ as  $J(\mathbf{r}) = \sum_{\mathbf{r'}} \theta(\mathbf{r'}) \theta(\mathbf{r'}-\mathbf{r})$, one obtains:
\begin{equation}
    \mathcal L_{\theta}(\hat{h},h) =\sum_{\mathbf{r},\mathbf{r'}} \big[\hat{h}(\mathbf{r})- h(\mathbf{r})]J(\mathbf{r'}-\mathbf{r})[\hat{h}(\mathbf{r'})- h(\mathbf{r'})\big].
    \label{LJ}
\end{equation}
Since the loss is essentially defined up to a constant, one can discard $\hat h$-independent terms in Eq.~\eqref{LJ}, such that  $\mathcal L_{\theta}(\hat{h},h) \equiv \mathcal H_h[\hat{h}]$ where:
\begin{multline}
    \mathcal H_h[\hat{h}] := \sum_{\mathbf{r},\mathbf{r'}} \hat{h}(\mathbf{r})J(\mathbf{r'}-\mathbf{r})\hat{h}(\mathbf{r'}) 
    - 2\sum_{\mathbf{r'}}(h \circledast  J)(\mathbf{r'})\hat{h}(\mathbf{r'}).
\end{multline}
Interestingly, when replacing $\hat h$ by a spin field $\phi$, one precisely recovers the Antiferromagnetic RFIM Hamiltonian $\mathcal H[\phi]$ which, within the bra-ket formalism writes:
\begin{equation}
    \mathcal H_h[\phi] =\langle \phi|J|\phi\rangle  - 2\langle h|J|\phi\rangle,
\end{equation}
where $|h_{\text{eff}} \rangle = 2J| h \rangle$ is the effective external field. Therefore, spins and external field in the Ising model respectively play the same role as the transformed and original images in the color quantization problem. A well-known property of the Antiferromagnetic RFIM, that is interactions encouraging alternations between neighbouring spins, is for example recovered in our observations, see Fig.~\ref{fig:2}(a).

\section{Resolution/relevance formalism} \label{Appendix:MIR}
Here we introduce the formalism used to set the optimal convolution parameter in Sec.~\ref{sec:}. The idea, detailed in \cite{cubero_minimum_2018,cubero_multiscale_2020,haimovici_criticality_2015}, consists in studying a data sample for different resolution/compression levels and calculating  entropy-based observables.

Let $\mathcal{S} =(s_1,...,s_N)$ be a sample of data, and $\ell$ a compression parameter. One can transform the original sample into a compressed one $\mathcal{S}^{\ell} =(s_1^{\ell},...,s_N^{\ell})$ such that $\ell=1$ corresponds to no compression ($\mathcal{S}^{1} = \mathcal{S}$) and $\ell = L$ to the totally compressed sample (quasi-degenerate data). At each compression level $\ell$, one can easily calculate the number $k_s^{\ell}$ of times the state $s$ appears in the sample $\mathcal{S}^{\ell}$, and the number $m_k^{\ell}$ of states appearing $k$ times. 
One can then define the \textit{resolution} as the entropy of the empirical distribution of states: 
\begin{equation}
    \hat H^{\ell}[s] = -\sum_s \frac{k^{\ell}_s}{N} \log\frac{k_s^{\ell}}{N} =-\sum_k \frac{k m_k^{\ell}}{N} \log\frac{k}{N}.
\end{equation}
Another useful observable is the \textit{relevance} defined as the entropy computed from the states sharing the same occurrence frequency:
\begin{equation}
\hat H^{\ell}[k] = \sum_k \frac{k m^{\ell}_k}{N} \log\frac{k m^{\ell}_k}{N}.
\end{equation}
This measure of \textit{relevance} is indeed the most direct way to encode the underlying distribution of the original data. {As the latter  entropy takes less bits to encode, one has $\hat H^{\ell}[k]<\hat H^{\ell}[s]$, explaining why the data in  Fig.~\ref{fig:MIR}(a) is under the $y=x$ line.}
In order to quantify the way information spreads across compression levels, we  define the compression rate $\mu_{\ell \to \ell +1}$ between two successive compression levels as the ratio between loss in relevance and loss in resolution:
\begin{equation}
    \mu_{\ell \to \ell +1} := \frac{\hat H^{\ell+1}[k]- \hat{H}^{\ell}[k]}{\hat{H}^{\ell+1}[s]-\hat{H}^{\ell}[s]}.
\end{equation}
In the right part of Fig.~\ref{fig:MIR}(a) and for the first compression steps (low values of $\ell$), $\hat H[k]$ is a decreasing function of $\hat H[s]$, resulting in negative $\mu$. Interestingly, this means that compressing the sample increases the amount of relevant information it contains about the underlying distribution.
However, when $\ell$ increases further, we reach an oversampling regime (left part of Fig.~\ref{fig:MIR}(a)) for which further compression reduces both relevance and resolution.

In the case of color quantization, we wish to gain information relevance as we range from low to higher compression in the most scale-invariant way possible. This way, we avoid the scale-specific tradeoff  described in Fig.~\ref{fig:3}. Optimality in that regard is therefore reached by taking $\mu$ close to -1 and as constant as possible in the undersampling regime, by that justifying the linear regression introduced in Sec.~\ref{sec:}.

\end{document}